\documentclass[12pt]{article}
\usepackage{amsmath,epsfig,graphicx,amssymb}

\newcommand{\beq}{\begin{equation}}
\newcommand{\eeq}{\end{equation}}
\newcommand{\ben}{\begin{eqnarray}}
\newcommand{\een}{\end{eqnarray}}
\newcommand{\bes}{\begin{subequations}}
\newcommand{\ees}{\end{subequations}}
\newcommand{\bFig}{\begin{figure}}
\newcommand{\eFig}{\end{figure}}

\date{}
\begin{document}

\title{Does Measurement Necessarily Destroy Coherence?}
\author{Partha Ghose\footnote{partha.ghose@gmail.com} \\
Centre for Astroparticle Physics and Space Science (CAPPS),\\Bose Institute, \\ Block EN, Sector V, Salt Lake, Kolkata 700 091, India}
\maketitle

\begin{abstract}
It has been proposed that measurement in quantum mechanics results from spontaneous breaking of a symmetry of the measuring apparatus and could be a unitary process that preserves coherence. Viewed in this manner, it is argued, non-destructive measurements should preserve this coherence and be reversible. It is shown that experiments with maximally entangled bipartite states can indeed distinguish between projective and unitary measurements.

\end{abstract}

\section{Introduction}
Measurement in quantum mechanics has been a source of much concern among physicists and philosophers \cite{genovese}. There are two dominant views about it, namely that of Bohr and von Neumann and their respective followers. According to Bohr, the measuring instrument must be a macroscopic object that behaves according to the rules of classical physics \cite{bohr1}. The measured system interacts with such an apparatus to form an {\em unanalyzable whole} which is mysterious. This makes the observed results apparatus-dependent or contextual and statistical in nature. In von Neumann's approach, the measuring apparatus is treated quantum mechanically, and the linear and unitary quantum evolution (`process 2') results in an {\em entanglement} of the system with the apparatus \cite{schr}. A measurement is then postulated as a non-unitary and non-quantum mechanical projection of this entangled state to one of the possibilities with a definite probability (`process 1') \cite{vN}. The origin of `process 1' is equally mysterious, and its necessity has been questioned \cite{ballentine}. It gives rise to the `measurement problem'. In both these views the unobserved system does not `possess' dynamical properties like position, momentum, spin components along specified directions, etc. though it can have definite kinematic properties such as mass, electric charge and other quantum numbers.

The description of unobserved systems as having no definite values of dynamical variables prior to measurement and the occurrence of definite values post-measurement is the hallmark of quantum mechanics that distinguishes it from classical physics. It is at the root of Bohr's Complementarity Principle. The requirement of Boolean-valued results post-measurement is natural in any branch of science in which unambiguous results are imperative. The non-existence of Boolean-valued variables in pure states pre-measurement and their resolution into mixed states with Boolean-valued variables post-measurement characterizes quantum measurements \cite{primas}.  

Irreversibility is another feature of measurements that is often emphasized. According to Wheeler's inerpretation of Bohr's view, for example, there is no `phenomenon' until it has been brought to a close by an irreversible act of amplification such as the blackening of a grain of silver bromide solution or the triggering of a photodetector \cite{wheeler}. In von Neumann's view, too, the non-unitary reduction or projection of a pure state to a mixed state by the act of measurement or observation is an irreversible process. There are, however, certain types of ideal von Neumann measurement known as `quantum nondemolition measurements' \cite{nondemo} that are irreversible but do not destroy the observed system and reproduce their outcome when repeated. `Null result measurements' are also possible in which a system need not be directly detected and demolished at all in order for a quantum measurement to occur -- the absence of detection and demolition can also provide definite information about a system \cite{renninger}. Irreversibility of such measurements is, however, not obvious and needs to be demonstrated \cite{ryff}.  

\section{Measurement as Spontaneous Symmetry\\ Breaking: Undoing a Measurement}
It has been recently proposed that measurement can be viewed as the result of spontaneous breaking of a symmetry of the measuring apparatus treated as a quantum mechanical system \cite{pg}. In this interpretation von Neumann's `process 1' is replaced by a unitary transition. Observations suggest that a quasi-closed system 
\beq
\vert X\rangle = \sum_{m, n} c_{m, n}\vert A_m \rangle \vert S_n\rangle  \label{x}
\eeq
in the Hilbert bundle describing the entangled system-measuring apparatus state becomes unstable against arbitrarily small environmental perturbations and spontaneously orients itself every time along one of the possible basis rays $\vert A_m\rangle \vert S_m\rangle$ with probability $\vert c_{m, m} \vert^2$:
\beq
M_{(m)} \vert X\rangle = \vert A_m\rangle \vert S_m\rangle
\eeq
with $M_{(m)}^\dagger M_{(m)} = 1$. {\em Since the ray orients itself along one of the basis rays every time, its projections on all the other basis rays vanish}. The environmental perturbations may be due to a larger system in which the system $\vert X\rangle$ is situated and/or due to vacuum fluctuations as in atomic transitions. The conventional projection is given by
\beq
\Pi_{(m)} \vert X\rangle = c_{m,m} \vert A_m\rangle \vert S_m\rangle
\eeq
with $\Pi_{(m)} = \vert S_m\rangle \vert A_m \rangle \langle S_m \vert\langle A_m \vert$. The other projections are set equal to zero by hand, resulting in the loss of information about the system $\vert X\rangle$. The principal difference from the conventional projection postulate is therefore simply this: whereas in the conventional case, a measurement result corresponds to a non-unitary and irreversible `projection' of the total ray to one of its possible basis rays followed by a normalization of the projected component (a patently non-quantum mechanical process), according to the new viewpoint the ray spontaneously orients itself completely along this ray with the same probability. Unitarity is preserved in the process, and the components along all other basis rays are automatically zero. These unitary transitions can be viewed as instances of spontaneous symmetry breaking (SSB) in the following sense. Let all states of the measuring apparatus (the ``pointer states'') before measurement be {\em a priori} equally probable. Then they can be viewed as degenerate ground states \cite{zimanyi}. The different possibilities given by the right-hand side of Eqn. (\ref{x}) can be viewed as ``attractors'' of the entangled state. When arbitrarily small and uniform perturbations are present, the state can make a unitary transition to one of these attractors, the choice of the particular attractor in any given event being by pure chance \cite{liu}. Once the state reaches one of these degenerate attractors, it becomes stable because there is no dynamical reason for it to shift to any other attractor.

Since the proposed measurement process results in the same outcomes with the same probabilities as a standard measurement in quantum mechanics, one may ask: is it yet another interpretation of quantum mechanics?  The answer is no. If measurement is indeed a unitary process, it must preserve coherence and be reversible, though this will however not be evident in cases where macroscopic irreversibility is involved. Let us consider a qubit state 
\beq
\vert \psi\rangle_0 = \alpha \vert 0\rangle + \beta \vert 1\rangle \label{1}
\eeq with $\vert \alpha \vert^2 + \vert \beta \vert^2 = 1$.
Let $M_0$ be a measurement such that 
\beq
M_0 \vert \psi\rangle_0 = \vert 0\rangle
\eeq and $M_1$ a measurement such that 
\beq
M_1 \vert \psi\rangle_0 = \vert 1\rangle.
\eeq In standard quantum mechanics these are assumed to be projections, and projections, being many to one, are irreversible. The standard projection operators $\Pi_i (i=0,1)$ for measurement on a qubit are defined by $\Pi_0 \Pi_1 = \Pi_1\Pi_0 = 0$ and $\Pi_0 + \Pi_1 = \mathbb{I}$. Let us represent the qubit state in the 2-dimensional space spanned by the basis vectors $\vert 0\rangle$ and $\vert 1\rangle$ by the column matrix
\[\vert \psi\rangle_0 =\left(\begin{array}{c}
 \alpha\\ \beta
\end{array} \right) \]
The projection operators in this space are reprented by the matrices

\[\Pi_0 = \left(\begin{array}{cc}
 1\,\,0 \\ 0 \,\, 0
\end{array} \right), \,\,\,\,\ \Pi_1 = \left(\begin{array}{cc}
 0\,\,0 \\ 0 \,\, 1
\end{array} \right)\]
which have no inverses, so that
\[\Pi_0 \vert \psi\rangle_0 = \alpha \left(\begin{array}{cc}
1 \\0 
\end{array} \right) = \alpha \vert 0\rangle, \,\,\,\,\ \Pi_1 \vert \psi\rangle_0 = \beta \left(\begin{array}{cc}
0 \\1
\end{array} \right)= \beta \vert 1\rangle\]
are irreversible processes.
The resulting states $\vert 0\rangle$ and $\vert 1\rangle$ have to be normalized by dividing by $\alpha$ and $\beta$ respectively for further calculations with them. 

On the other hand, it is possible to represent the measurements $M_0$ and $M_1$ by the {\em unitary} matrices

\[M_0 = \left(\begin{array}{cc}
 \,\,\,\,\alpha^*\,\,\,\,\,\,\, \beta^* \\-\beta \,\,\,\,\, \alpha
\end{array} \right), \,\,\,\,\ M_1 = \left(\begin{array}{cc}
\,\,\,\beta\,\,\,\,\,\,-\alpha \\\alpha^* \,\,\,\,\,\,\,\,\,\,\, \beta^* \end{array} \right)\]
so that
\[M_0 \vert \psi\rangle_0 = \left(\begin{array}{cc}
1 \\0 
\end{array} \right) = \vert 0\rangle, \,\,\,\,\ M_1 \vert \psi\rangle_0 = \left(\begin{array}{cc}
0 \\1
\end{array} \right)= \vert 1\rangle.\]
Notice that no further normalization is required as in the case of a projection. The inverses of these matrices are given by 
   
\[M_0^{-1} = M_0^\dagger = \left(\begin{array}{cc}
 \,\,\,\alpha\,\,\,\,\,\, -\beta^* \\ \beta \,\,\,\,\,\,\,\,\,\,\,\,\, \alpha^*
\end{array} \right), \,\,\,\,\ M_1^{-1} = M_1^\dagger =  \left(\begin{array}{cc}\beta^*\,\,\,\,\,\,\,\,\,\,\alpha \\-\alpha^* \,\,\,\,\,\,\, \beta \end{array} \right)\]
Notice that $M_0 M_1 \neq 0$, $M_1 M_0 \neq 0$ and $M_0 + M_1 \neq \mathbb{I}$, and hence $M_0$ and $M_1$ are not projection operators.
 
Now consider the density operator $\rho = \vert \psi\rangle_{00}\langle \psi\vert$ and 

\beq
\tilde{\rho}_0 = M_0 \rho M_0^\dagger
= \left(\begin{array}{cc}
1\,\,\,\, 0 \\ 0\,\,\,\, 0\end{array}\right),\,\,\,\,
\tilde{\rho}_1 = M_1 \rho M_1^\dagger = \left(\begin{array}{cc}
0\,\,\,\, 0 \\ 0\,\,\,\, 1\end{array}\right) 
\eeq Clearly, $\tilde{\rho}_i^2 = \tilde{\rho}_i (i=0,1), {\rm Tr} \tilde{\rho}_i = 1$, and the pure state remains a pure state in both the cases. On the other hand, it is well-known that
\beq
\tilde{\rho} = \Pi_0 \rho \Pi_0^\dagger + \Pi_1 \rho \Pi_1^\dagger  = \left(\begin{array}{cc}
\vert \alpha\vert^2\,\,\,\,\,\,\,\,\,\,\,\,\,\,\,\, 0 \\ \,\,\,\,0\,\,\,\,\,\,\,\,\,\,\,\,\,\,\,\,\,\, \vert \beta\vert^2\end{array}\right),\,\,\,\,\tilde{\rho}^2 \neq \tilde{\rho}, 
\eeq and one has to define the sub-ensemble 
\beq
\tilde{\rho}_\alpha = \frac{1}{\vert \alpha\vert^2}\left(\begin{array}{cc}
1\,\,\,\, 0 \\ 0\,\,\,\, 0\end{array}\right)
\eeq of the states $\vert 0\rangle$ by throwing away the sub-ensemble of the states $\vert 1\rangle$, and similarly 
\beq
\tilde{\rho}_\beta = \frac{1}{\vert \beta\vert^2}\left(\begin{array}{cc}
0\,\,\,\, 0 \\ 0\,\,\,\, 1\end{array}\right).
\eeq This results in the loss of information and non-unitarity.

Finally, let us consider the two operators
\beq
U = \left(\begin{array}{cc}
0\,\,\,\,\,\, e^{i\theta} \\ 1\,\,\,\,\,\,\, 0\end{array}\right), \,\,\,\, U^\dagger = \left(\begin{array}{cc}
0\,\,\,\,\,\,\,\, 1 \\ e^{-i\theta}\,\,\,\, 0\end{array}\right).
\eeq Let $\beta = \vert \beta \vert e^{i\phi}$. Then, straightforward calculations show that
\beq
U M_0 U^\dagger = M_0^{-1}, \,\,\,\,\,
U M_1 U^\dagger = M_1^{-1}\label{U1}
\eeq provided $\theta = - 2 \phi$.

If one chooses $\alpha$ and $\beta$ to be real and equal to $\frac{1}{\sqrt{2}}$, then it is easy to see that 
\ben
M_0 &=& M_1^{-1}= \frac{1}{\sqrt{2}}[\mathbb{I} + i\sigma_y]\label{4}\\
M_1 &=& M_0^{-1} = \frac{1}{\sqrt{2}}[\mathbb{I} - i\sigma_y]\label{5}
\een where $\sigma_y$ is a Pauli matrix, and
\beq
U = \left(\begin{array}{cc}
0\,\,\,\,\,\, 1 \\ 1\,\,\,\,\,\,\, 0\end{array}\right)\label{U2}
\eeq

\section{Qubits as Testing Grounds}

(i) Let us start with a heralded photonic state $\vert \psi\rangle_{in} = \vert 0\rangle$ and convert it into the qubit state $\vert \psi\rangle_0$ (Eqn. \ref{1}) with $\alpha = \beta = \frac{1}{\sqrt{2}}$ by passing it through a lossless $50-50$ beamsplitter: $M_1 \vert 0\rangle = \vert \psi\rangle_0$. Let $D_1$ be an ideal detector placed after the beamsplitter to detect the state $\vert 1\rangle$. Every time $D_1$ fails to detect a photon, a measurement $M_0$ occurs and the state $\vert 0\rangle$ results with $50 \%$ probability. Then the reversing operation $(M_0)^{-1} = M_1$ can be done with a second beamsplitter. The same final result would, however, hold if the first measurement were a projection $\Pi_0$. Hence, this type of experiment with a single qubit cannot distinguish between $\Pi_0$ and $M_0$. 

{\flushleft{(ii) Let us next consider the Bell state}}

\beq
\vert \Psi^+\rangle = \frac{1}{\sqrt{2}} [\vert 0\rangle_A \otimes \vert 1\rangle_B +  \vert 1\rangle_A \otimes \vert 0\rangle_B]\label{ent}
\eeq of two qubits labelled by $A$ and $B$. The Hilbert space is $H^{AB} = H^A\otimes H^B$ with the standard basis $\{\vert 0\rangle_A \otimes \vert 0\rangle_B, \vert 0\rangle_A \otimes \vert 1\rangle_B, \vert 1\rangle_A \otimes \vert 0\rangle_B, \vert 1\rangle_A \otimes \vert 1\rangle_B\}$ which we will henceforth label as $\{\vert 0 0\rangle, \vert 0 1\rangle, \vert 1 0\rangle, \vert 1 1\rangle\}$. Hence, the state (\ref{ent}) will be written as
\beq
\vert \Psi^+\rangle = \frac{1}{\sqrt{2}} [\vert 0 1\rangle +  \vert 1 0\rangle].\label{ent2}
\eeq
The important point about this bipartite state is that it can be written in terms of the two basis vectors $\vert 0 1\rangle$ and $\vert 1 0\rangle $ only. Hence, it can be written as a column vector in the 2-dimensional subspace of the full Hilbert space $H^{AB}$ spanned by $\vert 0 1\rangle$ and $\vert 1 0\rangle $:

\[\vert \Psi^+\rangle = \frac{1}{\sqrt{2}} \left(\begin{array}{cc}
1 \\1
\end{array} \right) \]
Now, define the unitary operators 
\[M^{AB}_0 = \frac{1}{\sqrt{2}}[\mathbb{I} + e^{i\pi/2}\sigma_y] = \frac{1}{\sqrt{2}}\left(\begin{array}{cc}
1\,\,\,\,\,\,\,\,\, 1 \\-1\,\,\,\,\,\, 1
\end{array} \right)\]
and

\[M^{AB}_1 =  \frac{1}{\sqrt{2}}[\mathbb{I} + e^{-i\pi/2}\sigma_y] = \frac{1}{\sqrt{2}}\left(\begin{array}{cc}
1\,\,-1 \\1\,\,\,\,\,\,\,\,\,\,1  \end{array} \right).\]
Their inverses are 
\ben
(M^{AB}_0)^{-1} &=& M^{AB}_1,\\
(M^{AB}_1)^{-1} &=& M^{AB}_0.
\een
It is easy to see that

\ben
M^{AB}_0 \vert \Psi^+\rangle &=& \left(\begin{array}{cc}
1 \\0
\end{array} \right) = \vert 0 1\rangle, \\
M^{AB}_1 \vert \Psi^+\rangle &=& \left(\begin{array}{cc}
0 \\1
\end{array} \right)= \vert 1 0\rangle,
\een and that

\ben
M^{AB}_1 M^{AB}_0 \vert \Psi^+\rangle &=& \vert \Psi^+\rangle,\\
M^{AB}_0 M^{AB}_1 \vert \Psi^+\rangle &=& \vert \Psi^+\rangle.
\een
In practice, one can use $U$, given by (\ref{U2}), as a flipper to implement the inverse operations $U M^{AB}_0 U^\dagger$ and $U M^{AB}_1 U^\dagger$ ((\ref{U1}), \cite{kim}).  

On the other hand, if measurements are projections $\Pi^{AB}_0$ and $\Pi^{AB}_1$, they would turn the Bell state $\rho = \vert\Psi^+\rangle\langle \Psi^+\vert$ into the mixed state $\tilde{\rho} = \sum_{i=0}^1\Pi_i\rho\Pi_i$, which will destroy its coherence irreversibly. Hence, a state like $\vert\Psi^+\rangle$ is a good candidate to use to test whether measurements are projective and irreversible or unitary and reversible. In practice it would be important to make use of `null result' measurements, and also to test whether the final state is a pure or mixed state by doing, for example, a quantum state tomography on it.

Similar analyses can be done in a straightforward manner for the other three Bell states $\{\vert \Psi^-\rangle, \Phi^{\pm}\rangle\}$ as well as for path-spin entangled states \cite{pathspin}. 

\section{Concluding Remarks}
I proposed earlier that a measurement occurs when the symmetry of the apparatus states is spontaneously broken, and is a unitary process. In this paper I have developed a simple mathematical formalism to describe unitary measurements in the case of single qubits and bipartite states such as Bell states and those involving path-spin entanglement. It is shown that non-destructive measurements on maximally entangled bipartite states like the Bell states can reveal to us whether measurements are fundamentally (a) non-unitary and irreversible, as assumed in standard quantum mechanics, or (b) unitary and reversible. Real experiments with such states should be of crucial importance in unravelling the true nature of measurement in quantum mechanics by telling us whether or not an `agent'/`observer' that is responsible for projective measurements, and hence not subject to quantum mechanics, is indeed indispensable. 

There already exists theoretical and experimental evidence that measurements can be undone, but such measurements are unsharp, {\em weak} and non-unitary \cite{kim, weak}. We are here concerned with strong and sharp measurements that are traditionally considered to be projective. 

After Ref. \cite{pg} appeared Vladan Plankovi\'{c} pointed out to me (private communication) that he and his collaborators have also proposed spontaneous symmetry breaking as a dynamical mechanism of superposition breaking \cite{planko}. Their proposal is, however, different from the one presented here in that in their scheme measurements are shown as continuous Landau phase transitions within a certain approximation. 

\section{Acknowledgement}
I would like to thank M. Genovese and Apoorva Patel for many helpful comments.
I also thank the National Academy of Sciences, India for the award of a Senior Scientist Platinum Jubilee Fellowship which allowed this work to be undertaken.

\end{document}